# M-GOVERNMENT, WILAYA AND WOMEN'S EMPOWERMENT IN THE KINGDOM OF SAUDI ARABIA

Norah Humus Alotaibi, The University of Sheffield, nhmalotaibi1@sheffield.ac.uk

Dr Salihu Dasuki, The University of Sheffield, s.dasuki@sheffield.ac.uk

Dr Efpraxia Zamani, The University of Sheffield, e.zamani@sheffield.ac.uk

**Abstract:** With the loosening of the male guardianship system (*Wilaya*) in the Kingdom of Saudi Arabia (KSA), this study examines the contribution of m-government to the empowerment of Saudi women. We employ the key concepts of Sen's Capability Approach (CA) to understand how m-government services in the KSA have provided opportunities for women to become empowered. The findings of the study indicate that m-government contributes towards women's empowerment by providing opportunities to participate in social and economic activities. However, there are also key social and cultural factors that impede the use of m-government services for empowerment, and we found these to include religious beliefs, Saudi traditions and customs, and husbands' jealousy. The study makes some important contributions to theory and practice by being the first study to focus on the use that Saudi women make of the opportunities now available to them to access government services through m-government applications and to address the cultural barriers which may function to prevent their access.

## 1. INTRODUCTION

With the advancement of Information and Communication Technology, governments are adopting electronic government (e-government) in order to provide efficient and effective government services to their employees, citizens, businesses and other organisations. However, with the rapid penetration of mobile technologies in everyday life, governments began shifting their focus towards the use of mobile phones for the delivery of their services and the provision of information to engage with citizens from anywhere at any time so that the latter can fully participate in political and social processes and reforms (Abu-Shanab & Al-Jamal, 2015; Lee et al., 2006).

There is large body of literature on e-government (Hsieh et al., 2013; Lee et al., 2011; Alawneh et al., 2013), however, studies focused on m-government are still very limited (Almaiah, Al-Khasawneh, Althunibat, & Khawatreh, 2020) in the Middle East and North Africa (MENA) context (Alharbi, Halikias, Yamin, & Basahel, 2020). The Kingdom of Saudi Arabia (KSA), the focal case of this study, is a highly influential country within the MENA region and has devoted a substantial amount of resources providing m-government services to its citizens (Alssbaiheen, & Love, 2015).

Yet there are only a few studies about m-government in the KSA, and the majority focus on m-government adoption and acceptance. In addition, scholars have argued that the adoption of m-government differs between men and women, with women usually being disadvantaged and having less access to e-government (Alzahran et al., 2018). M-government studies focusing on the KSA usually do not place an emphasis on women (Almarashdeh & Alsmadi, 2017; Alotaibi & Roussinov, 2017; Alonazi et al., 2019). Having said this, the KSA, which is a traditionally conservative Muslim country, places high value on male guardianship, female honour, male-female segregation and male-dominated practices, which continue to shape the use of m-government by women (Samargandi et al., 2019).

Today, there is a debate on whether m-government empowers or disempowers marginalised groups such as women (Gaur & Avison, 2015). There is a call for more studies to examine the experience of women using m-government, especially in the MENA region where the influence of Islam strongly shapes the gender roles (Ojediran & Anderson, 2020). To address this, our study asks: *How do women use m-government to empower themselves?* We interview women in the KSA, and examine the social, economic and political opportunities arising from their use of m-government services.

The KSA presents a compelling and unique case to examine the empowerment opportunities offered to women by the use of the m-government services. First, the KSA is a traditional Muslim country that struggles with equality for women due to religious doctrine. Second, the Saudi government has recently made reforms through its Vision 2030 to diversify the economy and encourage non-oil sectors with the goal of creating economic opportunities for women. Third, the government has loosened the restrictions placed on women by male guardianship in order to empower and elevate the status of women. We employ the Capability Approach as the theoretical lens as it argues for the freedom of women to lead the lives they value. To the best of our knowledge, our study is the first that investigates the contribution of m-government towards women's empowerment in Saudi Arabia, which is particularly important considering that the Saudi government has recently relaxed the *Wilaya* system, i.e., the male guardianship system that regulates the behaviour and movement of women, requiring them to seek the permission of a male relative for many of their activities. The study makes important theoretical and practical contributions for m-government studies and our findings have implications for policy makers and those who promote m-government services to empower Saudi women.

The remainder of this paper is structured as follows: the subsequent section presents a review of literature on women's empowerment and its relationship to m-government, followed by a discussion on the Capability Approach. The research method is then presented. We then discuss our case study and our findings, and finally, we offer a discussion of the theoretical and practical implications.

## 2. LITERATURE REVIEW

### 2.1 Women's Empowerment

Women's empowerment focuses on women having the opportunity to be able to pursue their choices and create social change (Pandey & Zheng, 2019). The empowerment of women is one of the United Nations' key Sustainable Development Goals (United Nations, 2015). However, there is still no clear pathway as to how gender empowerment can be achieved, possibly due to the lack of a clear conceptualisation of the term 'women's empowerment' (Nguyen & Chib, 2017). Some development scholars and activists view it from a perspective that focuses more on increasing women's access to employment opportunities, health care and education (Stavrou et al., 2015; United Nations, 2015) with less emphasis being placed on the agency of women to make choices in their lives (Nguyen & Chib, 2017). Recently, however, the concept of gender empowerment has been redefined, with greater emphasis placed on non-economic aspects and particularly on the agency of women (Donald et al., 2017).

This conceptualisation of women's empowerment resonates with Sen's (1999) Capability Approach (CA), which defines agency as "the freedom to achieve whatever the person, as a responsible agent, decides he or she should achieve" which is central to the process of development (1999, p. 18). In addition, Sen notes that "extensive reach of women's agency is one of the more neglected areas of development studies, and most urgently in need of correction" (ibid). In Information Systems (IS) studies, most development researchers have drawn upon the CA to conceptualise gender empowerment and show how technology as an artefact provides opportunities for women to create social change and pursue their choices (Pandey & Zheng, 2019).

Kumar, Karusala, Ismail and Tuli (2020) stress that measuring this empowerment is not purely a matter of looking at economic factors but at social welfare and how it is what women can do with the means offered to them (like m-government apps) that is key. These authors draw on ideas by Nussbaum (2011) to suggest that the agency of women in the Global South should take into account the severe gender inequalities of the region and include aspects of agency such as emotions related to the lack of freedom to express themselves without anxiety and fear (Kumar et al., 2020). In this study, empowerment, as conceptualised by Sen, serves as the theoretical basis to understand women's empowerment in a traditional, male-dominated, Islamic society. In the following section, we provide a review of the relationship between m-government and women's empowerment.

### 2.2 M-Government and Women's Empowerment

As developing countries continue to adopt e-government, many scholars have attempted to examine the developmental impact of e-government and most have shown that it has the potential to empower citizens by accessing information regarding various sectors of the economy (Hoque & Sorwar, 2015; Adaba & Rusu, 2014). This can improve people's ability to make a living, and ultimately, enhances human development.

Within developing contexts, telecentres are quite prominent and often used for accessing government services online. Taking a gender perspective, scholars have investigated their impact on women's empowerment. For example, Lwoga and Chigo's (2020) study in Tanzania shows that telecentres support women by increasing their income, saving money and voicing their concerns. Similarly, Hansson et al. (2010), show that telecentres in Sri-Lanka create new communication channels and career opportunities for women, and Alao et al. (2017) show that telecentres in rural South Africa empower women by enhancing their information capabilities, thereby improving their economic standards, their connections with friends and families, and ultimately their attainment of ICT literacy skills. Using mobile phones to access services potentially increases the ability to do this from anywhere. However, when considering the use of mobile phones for accessing such services in Ghana, Ojo et al. (2013) indicate that mobile phones alone are insufficient to provide women with opportunities or satisfy their information and livelihood needs as low levels of literacy are an important barrier.

Within the MENA region, and specifically the KSA, there is a paucity of research investigating the role of either e- or m-government for the empowerment of women. Saudi studies on m-government have focused mainly on the adoption or acceptance of m-government by both genders (Alrowili et al., 2015; Alotaibi & Roussinov, 2017; Alonazi et al., 2019). More importantly, scholars use quantitative methods, such as surveys, and focus on the single aspect of gender gap in m-government access. Such approaches do not help with understanding how women in this region are using m-government to improve their lives in light of significant social restrictions (Moussa & Seraphim, 2017).

In this study, we focus on women's voices to understand their experiences of using m-government services and to explore how such services contribute towards women's empowerment. In what follows, we present Sen's Capability Approach, which helps us understand the opportunities provided by government services for women to be empowered.

### 3. SEN'S CAPABILITY APPROACH

The CA framework is specifically suited to the context of our study due to its focus on social change in terms of enhancing individual wellbeing (Robeyns, 2016). The CA emphasises *human freedom,* which refers to the effective opportunities individuals have to enhance their well-being, and further critiques other development theories that emphasise wealth as central to happiness. The CA framework consists of two important elements: capabilities and functionings. Capabilities are the freedoms an individual has to attain a set of functions, whereas functionings are the "beings and doings" that an individual values (Sen, 1999, p. 18). The distinction between capabilities and functionings is between opportunities on the one hand and achievements on the other, respectively (Robeyns, 2005). However, Alkire (2005) noted that it is important to focus more on capabilities than functionings because individuals value having choices.

Commodities, such as goods and services, are a major constituent of the CA, because they are a means for enhancing individual wellbeing (Alkire & Deneulin, 2009). However, the ability to generate these capabilities is influenced by three sets of conversion factors, namely: personal conversion factors, e.g., gender, age, education; social conversion factors, e.g., social norms, public policies, power relations; and environmental conversion factors, e.g., infrastructure and climate (Robeyns, 2005). Achievements are based on an individual's choices from a set of capabilities, which is influenced by conversion factors (ibid). Another key feature of the CA is the difference between agency and wellbeing. Agency refers to the freedom to set and pursue one's own goals and interests (Sen, 1985), well-being relates to one's quality of life (Robeyns, 2005).

Within the area of Information Systems, and specifically the Information and Communication Technology for Development (ICT4D) domain, the CA has been applied as an evaluative framework to understand the contribution of ICT to human development (Zheng & Walsham, 2008), and has been applied within the context of e-government. For example, Adaba and Rusu (2014) used the CA to examine the impact of an e-government initiative for e-trade facilitation in Ghana. The study showed that the e-government initiative provided individuals and businesses with the opportunity and choice to lodge import and export declarations electronically which, in turn, enhanced the freedom of job creation. Also, Maiye and McGrath (2010) applied the concepts of CA to assess the Nigerian government's introduction of an electronic voter's registration system to enable participation in registration and polling activities. Their findings showed that the system expanded people's freedom to engage in development activities, but the sustainability of the developmental potentials were hindered by conversion factors such as lack of knowledge-building activities and lack of trust.

However, few studies have evaluated e-government initiatives using the CA framework from a gender perspective. Lwoga and Chigona (2020), for example, examined the use of telecentres by rural women in Tanzania, and showed that telecentres may enable women to build some capabilities (social, financial, human and political capabilities), but equally, they may inhibit others due to choices made and conversion factors, thus resulting in diverse development outcomes. Building on existing work by Lwoga and Chigona (2020); Maiye and McGrath (2010) *inter alia,* we apply the CA approach to examine the use of m-government services by women in Saudi Arabia.

## 4. METHODOLOGY

Following an interpretive approach, we conducted interviews with 30 Saudi women with the aim of understanding their use of m-government services and whether and how the use of such services contributes to their empowerment. The lead author, who is a Saudi citizen, used her personal contacts to recruit participants. Participants had to be (a) over eighteen years old (b) mobile phone owners and (c) m-government users. The interviews took place in January 2021, were conducted in Arabic and later translated to English. All transcribed data were uploaded to the Nvivo12 software package. Each interview lasted approximately 45 minutes and was conducted online due to Covid-19 restrictions. Prior to the interview, we sought informed consent, explaining the purposes of the research. To preserve confidentiality and anonymity, the names of the participants have been replaced with pseudonyms and any identifying information has been removed.

The CA framework guided the design of the interview questions, which were divided into three sections. The first covered demographic details; the second covered questions on participants' perceptions regarding empowerment, their use of m-government services and the opportunities derived from their use of m-government; and the final section focused on the factors affecting their use of m-government services.

After engaging with 30 participants, no new findings emerged from further probing and we stopped data collection, as the principle of data saturation had been satisfied. Data analysis for this study was influenced by the concepts of Sen's Capability Approach, following a thematic analysis (Braun & Clarke, 2006). We began by carefully reading the transcripts to identify any ambiguities and obtain a summary of the themes discussed by the women. Next, we merged these codes into broader categories. We particularly searched for, and identified, themes reflecting the m-government services used by the women, the opportunities provided to the women that contribute to their empowerment as a result of using the m-government services, and lastly, the factors affecting their use. In summary, our themes related to commodities (e-government services), the capabilities provided by the m-government services, and lastly, the conversion factors that influence their abilities to generate capabilities from the m-government services. Finally, the themes were reviewed and agreed upon by all the authors to ensure the analysis reflected the focus of the research. At the end of the analysis process, we produced our report as shown in the analysis section.

**Table 1 Participants' Demographic Information**

| Demographics | Frequency (n = 30) | Percentage (%) |
|---|---|---|
| **Age** | | |
| 19 – 25 | 3 | 10% |
| 26 – 35 | 9 | 30% |
| 36 – 45 | 12 | 40% |
| 46 – 55 | 4 | 13% |
| 56 – 65 | 1 | 3% |
| 66 + | 1 | 3% |
| **Education** | | |
| Bachelor | 18 | 27% |
| Masters | 2 | 7% |
| PhD | 2 | 7% |
| No education | 2 | 7% |
| Secondary education | 5 | 17% |
| British fellowship | 1 | 3% |
| **Marital Status** | | |
| Single | 9 | 30% |
| Divorced | 8 | 27% |
| Married | 10 | 33% |
| Widow | 3 | 10% |
| **Employment** | | |
| Housewife | 9 | 30% |
| Unemployed | 2 | 7% |
| Retired | 1 | 3% |
| Employed | 18 | 60% |

## 5. CASE STUDY CONTEXT

### 5.1 M-government in Saudi Arabia

Saudi Arabia recognises that ICTs support its economic growth. According to the National Transition 2020 and Saudi Vision 2030, ICTs are a key tool for economic development and for further diminishing the country's economic dependence on oil (Vision 2030, 2020). The government has started to transform all its traditional services into digital, in order to meet citizens' needs and expectations. According to Vision 2030, "Saudi Arabia has made remarkable progress in e-government over the last decade" (Vision 2030, 2020). The country has expanded its online services to include employment programmes, online job searches, e-learning services, traffic, passports and civil affairs, online payment services and online issuance of commercial registers. The rapid diffusion of mobile technologies has made the KSA the second biggest market in the Middle East for smartphones, with approximately 23 million smartphone users, allowing the transition from e-government to m-government in order to provide transparent, effective and convenient online channels between the government and citizens, accessible in a mobile environment (Alonazi & White, 2019). Currently, the KSA is making efforts towards facilitating a society built on knowledge, speed of response and interactivity, thus the quality and efficiency of m-government services are crucial (Gov. Saudi, 2020).

However, the adoption of Saudi m-government for citizens' empowerment has been problematic due to the slow adoption rate among citizens, particularly among women (Alghamdi & Beloff, 2016; Alotaibi et al., 2016). One of the main reasons for this is the male guardianship law that limits women from utilising government services that could enhance their social, economic and political wellbeing. However, with new reforms relaxing this law, it is pertinent to explore the impact of m-government on the empowerment of Saudi women and whether any improvements have been achieved. The following section provides an overview of the male guardianship system.

### 5.2 Wilaya – The Male Guardianship System in Saudi Arabia

All women in Saudi Arabia, irrespective of their age, are under the authority of a legal guardian who is a male relative and usually referred to as *Wali al-amr*. These male relatives, such as husbands, brothers and fathers, make an array of key decisions on behalf of any women under their watch. The women are also under the authority of other male relatives who have less control than their *Wali al-amr*. In addition, a male relative referred to as *mahram* (escort) has the authority to receive a woman who leaves her marriage due to domestic violence, or accompany her abroad on a government scholarship. Government institutions and agencies may ask a male relative to act as a woman's *mu'arif* –someone who is legally allowed to verify the identity of a woman covering her face, or carry out a range of important activities such as reporting a police case on her behalf and so on. A legal guardian can also act as a *mu'arif* or *mahram*. However, in 2019, the Saudi government legally enabled Saudi women to access government services as equal citizens and apply for their own government documents such as ID cards and passports, and allowed them to travel independently without the male guardianship system limiting their agency and freedom (Salaebing, 2019, Human Rights Watch, 2016).

## 6      FINDINGS AND ANALYSIS

We applied the concepts of commodities, capabilities and conversion factors as pillars for our analysis with the aim of examining the contribution of m-government services to Saudi women's empowerment. Absher and Yesser are the two m-government services used by our participants, and following the Capability Approach, these two services can be seen as the commodities that may help women enhance their status and lead to their empowerment. In other words, m-government services are resources which allow Saudi women to choose things (such as passports, driving licences ID cards etc.,) which gives them the freedom to do the things they value (such as travelling abroad to gain academic qualifications), and the extent to which they are empowered to do this depends on 'conversion factors' (such as the norms and values held by their family about such matters).

### 6.1   Commodities

In 2005, a programme called 'Yesser' was launched with the aim of developing e-services that would link up the different ministries and ensure that all Saudi government agencies had their own websites (Alshehri & Drew, 2010). Since then, many ministries have launched websites, such as the Ministry of Health (www.moh.gov.sa), the Ministry of Labour (www.mol.gov.sa) and the Ministry of Education (www.moe.gov.sa).

'Yesser' in Arabic means 'make it easy', and the programme aims to provide services and information that citizens can access easily (Alghamdiv, 2016). The e-government programme was launched by the Ministry of Communication and Information Technology (MCIT) in partnership with the Ministry of Finance and the Communication and Information Technology Commission (CITC). The overarching aim of Yesser is to reduce the digital divide in the Saudi community and transform the country into a digital society by delivering information through electronic channels and to improve response times among government services (Yesser, 2006).

The Absher app is the official mobile application that delivers government services to all citizens and residents. It is available in both Arabic and English and has been designed with special consideration for the security and privacy of user data. Users can safely browse and update/renew visas for their employees, and perform a range of services online. It can be used for job seeking, applying to attend the annual *Hajj* Muslim pilgrimage, updating passport information and reporting electronic crimes. In addition, it provides a gateway to 160 services, including making appointments, renewing passports, applying for residency cards, IDs, driving licenses and others. Currently, Absher has more than 11 million users across the KSA (MOI, 2020).

## 6.2 Capabilities

### 6.2.1 Opportunities for Mobility and Economic Activity

Using the Absher system, women are able to apply for or renew their passports without the permission of a male guardian. The ability to acquire a passport independently provides women with freedom of movement, making it possible for them to travel within and outside the country, pursue academic studies or attend business conferences. As one divorced, employed, participant noted:

*"I was never able to travel abroad with my kids because my male guardian was totally against it and would not sign the documents. Now, I just applied and obtained a new passport on Absher for myself and the children and immediately the lockdown is eased we are taking a holiday in Dubai."*

In addition, women can access Absher to apply for a driver's license which further enhances their economic freedom. Many women can drive for the first time and even own a car in their own name. Participants noted that being able to drive saves them the cost of using public transportation. In the past, women relied on their male guardians or hired foreign male drivers, which could be very expensive. Women who could not afford a chauffeur often had to give up on economic opportunities outside their home, as noted by this participant:

*I sell traditional beauty products and before I could not always get to my clients due to the high cost of transportation and having to seek permission every time from my male guardian. However, now I have officially registered my business on the Ministry of Commerce website, without needing anybody's permission, obtained a driving license and bought a car. So now I can save money and easily visit customers and attend business conferences."*

Obtaining a driver's license means that women can register on government websites as taxi drivers and provide their services to other women clients. Thus, they now have the opportunity to generate income without depending on others. One of the participants who did not have a degree and was previously unable to get a job said:

*"Thank God for Absher. I am a divorcee with no source of income and depended on my male guardian. Sometimes he did not provide the financial support I needed. With Absher, I was able to get a driving license and get a car loan, so now I work as a taxi driver and earn money to support myself."*

Lastly, under the new system, women now have the opportunity to upload their CVs to Absher and seek employment through the government.

### 6.2.2  Opportunities for Self-Identification

The Absher system enables women to register and obtain a national ID card online. This in turn enables them to carry various necessary activities independently. This was not possible in the past, as women could only be registered under their male guardian's national ID card. The new system allows women to register births and deaths, open bank accounts and more importantly register their own children and enrol them in school. According to one of the participants, having a national ID card and the ability to access it via their mobile phones helped solved a lot of problems associated with male guardianship:

*"Before, there was nothing to prove that they were my children, but now there is a new family record attached to my ID card, so I can change my children's school easily. Because registration was in the name of the male guardian, previously only their father could apply to schools, but now the family record allows me to apply to the schools that I want for my children."*

Some divorced participants mentioned that having the ability to obtain a national ID card with information about their children stopped their ex-husbands refusing to enrol their child in school as a form of punishment for the women. Another widowed participant noted how she was now able to put money in the bank for her children:

*"Under the old system, the children's uncle would be their Wali and be responsible for them. I could not interfere with this. I mean, even school registration, applying for their papers, opening their bank accounts and registering them for university had to be done by their uncle. With the new change, I am now solely responsible for my children. Before my husband died, he was in a coma, and it was before the decision to allow a woman to act without a guardian. I went to deposit an amount in my older son's account. However, the bank told me I was forbidden to deposit something in my children's accounts without the written consent of their father. So what about after he died? How difficult that situation was for me! Now I am responsible for them - not their uncle."*

Using Absher allows women to book hospital appointments for themselves and their children and also access their medical records and provide consent where necessary. Further, it provides support for online consultations with doctors, which was especially beneficial during the pandemic:

*"The Ministry of Health has just announced a 24/7 hotline to answer questions about various diseases which is an excellent step and was very helpful to me personally because I cannot leave my home for prolonged periods, so viewing medical instruction videos and having online consultations comes in handy."*

### 6.2.3 Opportunities for Better Education and Voicing Concerns

Some participants mentioned that in the past their male guardians refused to give consent or sign their documents to allow them go abroad to further their education. However, through Absher, one participant, who is a lecturer, was able to apply and secure a PhD scholarship without such obstacles:

*"I wanted to obtain a PhD in order to gain promotion at the university. But my uncle kept refusing to give consent, stating I must marry first. However, with the new system, I simply applied and I will start my PhD in October. It has always been my dream to be called 'doctor'."*

Another participant noted that her male guardian kept taking advantage of his authority over her to extort money from her as a condition for allowing her study in the university. However, the new system has now enabled her to secure a university admission and begin her studies:

*"Two years ago, my uncle, who is against women having a formal education, asked me for a large sum of money as a condition of allowing me to enrol for a degree, but he never approved or signed*

*my documents. But now, with the new system I have been able to start my degree and will be finishing in 2024, God willing."*

The Absher system further enables women to report their concerns and inappropriate behaviour. For example, one of the women explained how she was able to report, via Absher, a man who directed abuse to her while she was driving:

*"I was driving when a male driver insulted me and called me names because I was driving. I immediately took a video and his number plate and reported the case. The case was dealt with and he was fined and asked to make a formal apology to me".*

Others mentioned that the system allowed them to report cases of discrimination towards them by institutions as well as individuals hence empowering them to voice their concerns about things that mattered to them.

### 6.2 Conversion Factors

The participants' ability to generate capabilities is influenced by conversion factors. The analysis of the empirical material shows that the relaxation of the male guardianship law (social conversion factor) enables women to generate capabilities using m-government services. However, there are other social conversion factors that may also hinder women from using m-government services towards building their capabilities.

For example, some of the women reported that it was against their religious values to use m-government services:

*"Not only because using Absher is against Saudi customs and traditions, but because it isn't permissible from a religious perspective as it involves mingling with men, not to mention the fact that you have to show your face for the photo. Women like me believe that a woman exposing her face is against the rules. Absher involves me taking off my Niqab for the photo and exposing my face for men to see."*

Other women noted that they respect tradition and still hold a strong belief in male guardianship. They further noted that a woman should respect her male guardian and seek his permission and that it is the responsibility of men to carry out these activities. Others mentioned the issue of jealousy from their partners who believe that a woman should always be under the authority of a man:

*"My husband refuses to let me take up a scholarship as it means travelling while he is at work and he would not let me go out alone because of jealousy. He feels other men might see me or make advances to me. He still believes in the male guardianship law and despite the law reform he made it clear I could not go. I cannot apply for a scholarship or request a driver's license in case he checks my phone. My husband is very jealous."*

### Table 2 Summary of Findings

| Commodities | Conversion Factors | Capabilities |
|---|---|---|
| M-Government Platforms<br>• Absher<br>• Yasser | Religious Beliefs | Mobility |
| | Traditions and Customs | Economic activities |
| | Jealousy | Self-identification |
| | Social (relaxation of male Guardianship) | Responsibility for own children/Childcare |
| | | Better health care |
| | | Education |

| | | Voicing their concerns |
|---|---|---|

# 7    CONCLUSION

The CA used in this study is part of the human development approach which emphasises that the aim of human development is to enhance peoples' lives by allowing them to access a variety of choices, such as being healthy, being knowledgeable and participating in their community. Such development means aiming to eradicate the barriers that restrict peoples' agency, like illness, illiteracy, lack of civil freedoms and poor access to resources. This framework emphasises that social change should encourage peoples' capability and provide them with social, economic and political freedom to choose and enjoy their own lives and what they are and do (Alkire & Deneulin, 2009).

In the context of this study, the obstacle was lack of access to resources, as Saudi women were not allowed to access m-government services without male permission before 2019. This means that the previous system, which only allowed males to directly access m-government and controlled women's access, reduced Saudi women's quality of life. Legally, Saudi women now have the freedom to access government services as equal citizens to apply for their own government documents such as ID cards and passports, as they can travel independently without the male guardianship system limiting their agency and freedom.

The study findings show that the use of m-government services enhances the capabilities of women to attain empowerment. These capabilities are related to mobility, economic activities, self-identification and responsibility for children, health care, education and having a voice. Using the CA, our study shows that accessing m-government services contributed to the freedom of women to participate in social and economic activities that could lead to their empowerment. However, the abilities to generate capabilities from the m-government services were hindered by social conversin factors such as religious belief, tradition and customs, belief in male guardianship, and lastly, male jealousy. In sum, while m-government services have improved the capabilities of women in the KSA, what remains to be seen is whether women will be fully empowered, considering the strong Islamic beliefs and traditions that continue to shape the way of life.

As a contribution, the study adds to the literature on ICT4D and women's empowerment by providing insights into the use of m-government by Saudi women. To the best of our knowledge, no study in the field has been conducted in the Saudi context. In proposing areas for potential ICT4D research, the limitation of this study is acknowledged. The study was carried out under severe time constraint with a small sample size. However, there is scope for conducting a longitudinal study. Future studies could further investigate how conversion factors also influence the intention of women to use m-government for their empowerment.

Although care was taken to make sure that the sample of participants represented Saudi women from a range of backgrounds, economic and marital statuses and educational levels, there were no systematic comparisons made between these groups. In order to explore how Saudi women from different socio-economic groups experience empowerment afforded by m-government and how the barriers to that empowerment (such as the Wilaya system) operate within these groups, the authors would recommend further comparative research with samples stratified by specific socio-economic factors such as income, occupation, marital status and so on to identify their influence on empowerment.

Alotaibi et al.                                    M-government, Wilaya, and Women Empowerment in Saudi Arabia
Alao, A., Chigona, W., & Lwoga, E. T. (2017). Telecentres bridging digital divide of women in rural areas : Case of Western in Partnership for Progress on the Digital Divide 2017 International Conference, 24-26 May 2017, Best Western Plus Island Palms Hotel & Marina San Diego, California USA

Alawneh, A., Al-Refai, H., & Batiha, K. (2013). Measuring user satisfaction from e-Government services: Lessons from Jordan. Government Information Quarterly, 30(3), 277–288. https://doi.org/10.1016/j.giq.2013.03.001

Alharbi, A. S., Halikias, G., Yamin, M., & Basahel, A. (2020). An overview of M-government services in Saudi Arabia. International Journal of Information Technology (Singapore), 12(4), 1237–1241. https://doi.org/10.1007/s41870-020-00433-9

Alkire, S. (2005). Why the Capability Approach? Journal of Human Development, 6(1), 115–135. https://doi.org/10.1080/146498805200034275

Alkire, S., & Deneulin, S. (2009). The human development and capability approach. In S. Deneulin (Ed.), An introduction to the human development and capability approach. London, England: Earthscan.

Almaiah, M. A., Al-Khasawneh, A., Althunibat, A., & Khawatreh, S. (2020). Mobile Government Adoption Model Based on Combining GAM and UTAUT to Explain Factors According to Adoption of Mobile Government Services. International Journal of Interactive Mobile Technologies (IJIM), 14(03), 199. https://doi.org/10.3991/ijim.v14i03.11264

Almarashdeh, I., & Alsmadi, M. K. (2017). How to make them use it? Citizens acceptance of M-government. Applied Computing and Informatics, 13(2), 194–199. https://doi.org/10.1016/j.aci.2017.04.001

Alonazi, M., Beloff, N., & White, M. (2019). Developing a Model and Validating an Instrument for Measuring the Adoption and Utilisation of Mobile Government Services Adoption in Saudi Arabia. Proceedings of the 2019 Federated Conference on Computer Science and Information Systems, 18, 633–637. https://doi.org/10.15439/2019f43

Alotaibi, R., Houghton, L., & Sandhu, K. (2016). Exploring the Potential Factors Influencing the Adoption of M-Government Services in Saudi Arabia: A Qualitative Analysis. International Journal of Business and Management, 11(8), 56. https://doi.org/10.5539/ijbm.v11n8p56

Alotaibi, S., & Roussinov, D. (2017). User acceptance of M-government services in Saudi Arabia: An SEM approach. Proceedings of the European Conference on E-Government, ECEG, Part F1294, 10–19.

Alrowili, T. F., Alotaibi, M. B., & Alharbi, M. S. (2015). Predicting citizens' acceptance of M-government services in Saudi Arabia an empirical investigation. 9th Annual IEEE International Systems Conference, SysCon 2015 - Proceedings, 627–633. https://doi.org/10.1109/SYSCON.2015.7116821

Alshehri, M., & Drew, S. (2010). Challenges of e-government services adoption in Saudi Arabia from an e-ready citizen perspective. World Academy of Science, Engineering and Technology, 66, 1053–1059.

Alssbaiheen, A., & Love, S. (2015). Exploring the Challenges Of M-Government Adoption in Saudi Arabia. Electronic Journal of E-Government, 13(1), 18–27.

Alzahrani, L., Al-Karaghouli, W., & Weerakkody, V. (2018). Investigating the impact of citizens' trust toward the successful adoption of e-government: A multigroup analysis of gender, age, and internet experience. Information Systems Management, 35(2), 124–146. https://doi.org/10.1080/10580530.2018.1440730

Awotwi, J., Ojo, A., & Janowski, T. (2011). Mobile governance for development - Strategies for migrant head porters in Ghana. ACM International Conference Proceeding Series, 175–184. https://doi.org/10.1145/2072069.2072099

Braun, V., & Clarke, V. (2006). Using thematic analysis in psychology. Qualitative Research in Psychology, 3(2), 77–101. https://doi.org/10.1191/1478088706qp063oa
Proceedings of the 1st Virtual Conference on Implications of Information and Digital Technologies for Development, 2021

497